\documentclass[prl,twocolumn,superscriptaddress,floats,showpacs]{revtex4}

 \usepackage[dvips]{graphicx}
 \usepackage[dvips]{graphics}

\usepackage[dvips]{graphicx}

\begin{document}
\title{Evidence for Intra-Unit-Cell magnetic order in  $\rm Bi_2Sr_2CaCu_2O_{8+\delta}$}

\author{S. De Almeida-Didry}
\affiliation{Laboratoire L\'eon Brillouin, CEA-CNRS, CEA-Saclay, 91191 Gif sur Yvette, France}
\affiliation{GREMAN, UMR CNRS 7347, Universit\'e de Tours, IUT de Blois, 15 rue de la chocolaterie, CS2903, 41000 Blois, France}

\author{Y. Sidis}
\affiliation{Laboratoire L\'eon Brillouin, CEA-CNRS, CEA-Saclay, 91191 Gif sur Yvette, France}

\author{V. Bal\'edent}
\affiliation{Laboratoire L\'eon Brillouin, CEA-CNRS, CEA-Saclay, 91191 Gif sur Yvette, France}

\author{F. Giovannelli}
\affiliation{GREMAN, UMR CNRS 7347, Universit\'e de Tours, IUT de Blois, 15 rue de la chocolaterie, CS2903, 41000 Blois, France}

\author{I. Monot-Laffez}
\affiliation{GREMAN, UMR CNRS 7347, Universit\'e de Tours, IUT de Blois, 15 rue de la chocolaterie, CS2903, 41000 Blois, France}

\author{P. Bourges}
\email{philippe.bourges@cea.fr} \affiliation{Laboratoire L\'eon Brillouin, CEA-CNRS, CEA-Saclay, 91191 Gif sur Yvette, France}

\date{\today}


\begin{abstract}
Polarized elastic neutron scattering measurements have been performed in the bilayer copper oxide system $\rm Bi_2Sr_2CaCu_2O_{8+\delta}$, providing evidence for an intra unit cell (IUC) magnetic order inside the pseudo-gap state. That shows time reversal symmetry breaking in that state as already reported in $\rm Bi_2Sr_2CaCu_2O_{8+\delta}$ through dichroism in circularly polarized photoemission experiments. The magnetic order displays the same characteristic features as the one previously reported for monolayer $\rm HgBa_2CuO_{4+\delta}$ and bilayer $\rm YBa_2Cu_3O_{6+x}$, demonstrating that this genuine phase is ubiquitous of the pseudo-gap of high temperature copper oxide materials. 
\end{abstract}

\maketitle



The existence of the mysterious pseudo-gap (PG) state in the phase diagram of copper oxide superconductor and its interplay with unconventional {\it d-wave} superconductivity has been a long standing issue for more than a decade. There is now a growing number of experimental indications that the pseudo-gap phase actually corresponds to a symmetry breaking state \cite{Kaminski-Dichroism,CC-review,Kapitulnik,Lawler,Pb-Bi2201}. In his theory for cuprates,  C. M. Varma \cite{Varma97,Varma-Simon} proposes that PG is a new state of matter associated with the spontaneous appearance of circulating current (CC) loops within $\rm CuO_2$ unit cell. This Intra-Unit-Cell (IUC) order breaks time reversal symmetry, but preserves lattice translation invariance. While from a theoretical point of view the existence of a CC-loop order and the ability of such a q=0 instability to produce a gap in the charge excitation spectrum are still highly controversial \cite{Greiter07,Weber,Gabay08}, several experimental observations provide strong encouragement for models based on CC-loop order in copper oxide materials. Polarized elastic neutron scattering studies carried in bilayer $\rm YBa_2Cu_3O_{6+x}$ (Y123)  \cite{Fauque,Sidis,Mook,Baledent-YBCO} and monolayer $\rm HgBa_2CuO_{4+\delta}$ (Hg1201) \cite{Li-Nature,Li-PRB} have reported experimental evidence of a long range 3D magnetic order hidden in the PG state. In $\rm La_{2-x}Sr_xCuO_4$ (La124), a similar magnetic order has also been observed \cite{Baledent-LSCO}, but in this system it remains 2D and short ranged. This novel magnetic state preserves the lattice translation invariant, but, at variance with ferromagnets, does not give rise to a uniform magnetization \cite{CC-review}. These observations imply the existence of an IUC (antiferro-)magnetic order, whose symmetry is consistent with the  so-called CC-$\rm \theta_{II}$ phase proposed by C. M. Varma \cite{Varma97,Varma-Simon}. Within that model, neutron diffraction measures the distribution of static magnetic fields generated by the CC-loops. The possible detection of these magnetic fields by local probes is still debated \cite{Lederer,Strassle-2}, but should help to get a deeper understanding of the instrinsic nature of the IUC magnetic order.  In Y123, the observation of anomalies in the  second derivative of the magnetization \cite{Leridon} when the IUC magnetic order settles in and the breaking of time reversal symmetry  below the ordering temperature in Kerr effect measurements \cite{Kapitulnik} can also be understood within the framework of the CC-loop model \cite{Gronsleth,Aji}, putting additional symmetry constraints\cite{Orenstein}. In addition, observation of CC-loops in $\rm CuO_2$ unit cell has been recently reported in non superconducting CuO materials using resonant X-ray diffraction \cite{Scagnoli}.

Here, we report a polarized neutron diffraction study of the bilayer cuprate, $\rm Bi_2Sr_2CaCuO_{8+\delta}$ (Bi2212). Up to now, the most accurate information concerning the electronic properties in the PG state have been provided through tunneling spectroscopy \cite{Renner-RMP} and angle resolved photo-emission (ARPES) \cite{Damascelli-RMP} carried out mostly in this system. Our study shows that the long range IUC magnetic order found in Hg1201 and Y123 is also present in Bi2212. This order develops at a temperature $\rm T_{mag}$ close to the PG temperature $\rm T^{\star}$ reported by various techniques. The observation of an IUC magnetic order in Bi2212 confirms that time reversal symmetry is broken in the PG state, as first suggested by the circular dichroism in ARPES \cite{Kaminski-Dichroism,Varma-Simon,Agtenberg}. Likewise, our polarized neutron diffraction study in Bi2212 and the analysis of the STM images \cite{Lawler} in the same system suggest that IUC order is  likely involved in the PG physics.

\begin{figure}[t]
\includegraphics[width=6cm,angle=0]{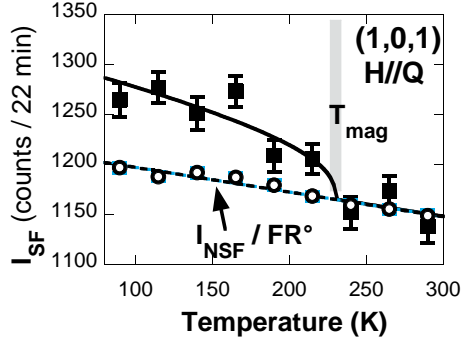}
\caption { Temperature dependence of the intrinsic Bragg scattering at {\bf Q}=(1,0,1) in the SF channel (full squares) and of polarization leakage (opened circles) corresponding to a T-independent bare flipping ratio $\rm FR^o$= 52.6. }
\label{Fig1}
\end{figure}

Measurements are performed on underdoped (UD) $\rm Bi_2Sr_2CaCu_2O_{8+\delta}$ (Bi2212) single crystals. The synthesis is carried out using the travelling solvent floating zone technique (TSFZ) in air \cite{DeAlmeida-sample}. Three large  crystals have been extracted from the as-grown rod. The composition homogeneity and the bulk crystal quality is provided by EDX and neutron diffraction studies. The as-grown singles crystals are weakly overdoped (OD) with a nominal superconducting critical temperature $\rm T_c$ of 87 K. The underdoping of the samples is then achieved using a post-annealing treatment of 300 hours under reduced oxygen atmosphere $\rm P(O_2)$=0.05 atm at 450$^{o}$C, yielding an average onset $\rm T_c$ of 85 K \cite{DeAlmeida-sample}. 

For the neutron diffraction measurements, the co-aligned single crystals are attached on the cold head of a 4K-closed cycle refrigerator and aligned in the [100]/[001] scattering plane, so that tranferred wave vectors $ \rm \bf{Q}$ of the form (H,0,L) are accessible. $ \rm \bf{Q}$ is given in reduced lattice units $ \rm  ( \frac{2 \pi}{a}, \frac{2 \pi}{b}, \frac{2 \pi}{c}) $, using tetragonal notations $\rm a \simeq b=$ 3.82 ~\AA  ~and c=30.87 ~\AA. Polarized neutron diffraction measurements are performed on triple-axis spectrometer 4F1 at reactor Orph\'ee in Saclay (France). The polarized neutron scattering set-up is similar to the one used in previous experiments on the same topic \cite{Fauque,Sidis,Mook,Baledent-YBCO,Li-Nature,Li-PRB,Baledent-LSCO,CC-review}. The scattered intensity at a given wave vector $\bf{Q}$ is measured in the spin-flip (SF) and non-spin-flip (NSF) channels, with the neutron spin polarization  $\rm \bf{H} // \bf{Q}$. For this polarization, the full magnetic intensity always appears in the SF channel. Due to a large neutron depolarizion when entering the SC state in Bi2212 samples, the identification of any magnetic signal below $\rm T_c$ is prohibited using our current neutron polarization 
device \cite{CC-review}. 

Following previous studies in bilayer Y123 \cite{Fauque,Sidis,Mook,Baledent-YBCO} and Hg1201 \cite{Li-Nature,Li-PRB}, the search for a long range magnetic order in the PG phase is performed on Bragg reflections (1,0,L)-(0,1,L) with integer L values.
The scattered intensity in the SF channel on a Bragg reflection ($\rm I_{SF}$) is dominated by the leakage of the NSF intensity  into the SF channel, whose magnitude gives the bare flipping ratio ($\rm FR^o(T)$), characterizing the neutron beam polarization quality and stability. On top of this signal, the intrinsic magnetic response ($\rm I_{mag}$), of much weaker intensity, can develop once a magnetic order settles in below a certain temperature. The scattered intensity in the SF channel then reads:
\begin{equation}
I_{SF}=  I_{NSF} / FR^o(T) + I_{mag}
\label{I-SF}
\end{equation}
As previosuly discussed\cite{Baledent-YBCO}, the neutron intensity $I$ stands in Eq. \ref{I-SF} for the intrinsic Bragg intensity, {\it i.e.} the intensity measured at the Bragg peak to which a background is removed. The background is typically measured away from the Bragg position, for instance at Q=(0.9,0,L) for Q=(1,0,L).

The averaged crystal structure of Bi2212 is usually described by a Bb2b space group: in our tetragonal notations, (H,0,L) Bragg reflections are thus observable {\it only} for H+L even. We then first search for magnetic scattering at Q=(1,0,L) for odd L values. Fig.~\ref{Fig1} shows the SF and NSF intensities for $\rm \bf{Q}$=(1,0,1) where an enhancement of $\rm I_{SF}$ occurs below $\sim$ 230 K indicating the appearance of a magnetic order in our underdoped Bi2212 sample.
The same kind of magnetic signal appears below $\sim$230 K at L=1 and L=3. Increasing further L to 5 and 7, the magnetic signal vanishes. Searches for the existence of a magnetic signal at even integer L values (L=0,2) or for non integer L have remained unsuccessful. This study of the magnetic scattering along (10L) emphasizes that the magnetic order is likely resolution-limited, {\it i.e.} long range. It should be noted that Bi2212 actually exhibits an incommensurate crystal structure with very strong satellite reflections\cite{Etrillard}. That precludes the determination of the absolute values of the magnetic moments as nuclear Bragg scattering intensities cannot be estimated with enough accuracy. 

\begin{figure}[t]
\includegraphics[width=8cm,angle=0]{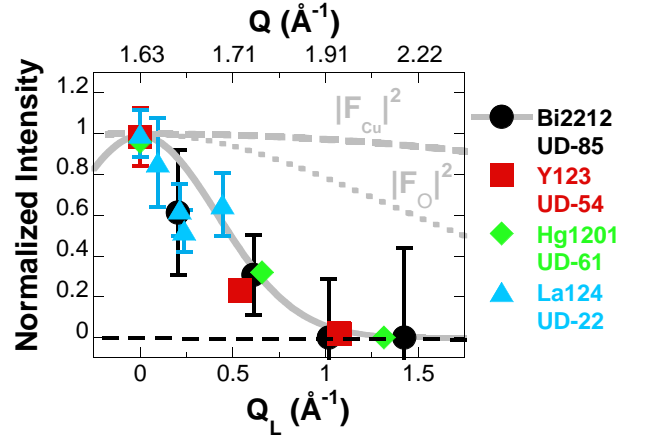}
\caption {
(color online) Variation of the magnetic signal measured above $\rm T_c$ at $ \rm \bf{Q}$=(1,0,L) as a function of the wave evector perpendicular to the ${\rm CuO_2}$ plane: $\rm Q_L=\frac{2 \pi} {c} L$ for different cuprates:  Hg1201 (UD-61) \cite{Li-Nature}, Y123 (UD-54)\cite{Sidis}, La124 (UD-22)\cite{Baledent-LSCO} and Bi2212 (UD-85) (see text for scaling procedure). The upper horizontal scale indicates the full momentum $\rm Q=\sqrt{(\frac{2 \pi} {a} H)^2+(\frac{2 \pi} {c} L)^2}$.}
\label{Fig2}
\end{figure}

As in monolayer Hg1201 \cite{Li-Nature,Li-PRB}  and bilayer Y123  \cite{Fauque,Sidis,Mook,Baledent-YBCO}, a 3D magnetic order then develops in the underdoped phase of bilayer Bi2212. In  monolayer La124 \cite{Baledent-LSCO}, this magnetic order remains 2D and the magnetic scattering spreads along the (1,0,L) rod.  The magnetic intensities along the $\rm {\bf c}$ axis are reported on Fig.\ref{Fig2} for these 4 cuprates families. They have been all rescaled to unity for L=0 (In Bi2212, they are actually normalized to 0.6 at L=1, $\rm Q_L$=0.20~\AA$\rm ^{-1}$, as the (1,0,0) peak is absent). Interestingly, $\rm I_{mag}$ exhibits a similar decay with $\rm Q_L$ for all systems suggesting a common magnetic origin. It is worth to mention that $\rm I_{mag}$ decays much faster than the squared  magnetic form factor of Cu$\rm ^{2+}$ and O$\rm ^{2-}$ ions (reported in Fig.~\ref{Fig2} for a sake of comparison).

\begin{figure}[t]
\includegraphics[width=7cm,height=8cm,angle=0]{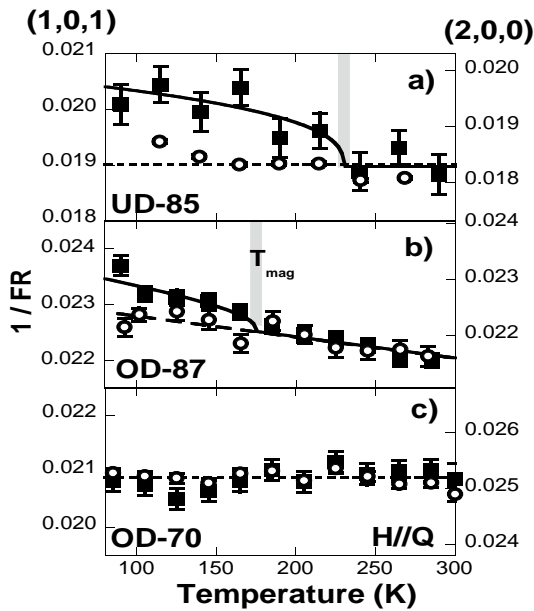}
\caption {Inverse flipping ratio, $\rm 1/FR(T)$, defined as Bragg intensity in SF channel normalized by the Bragg intensity in the NSF channel: {\bf Q}=(1,0,1) (full squares) and {\bf Q}=(2,0,0) (open circles) for 3 samples: a) UD-85, b) OD-87 , c) OD-70.  }
\label{Fig3}
\end{figure}

We now focus on the hole doping dependence of the magnetic signal through a comparative study with  2 other overdoped (OD) single crystals: OD-87 \cite{Fauque-Bi2212} and OD-70 \cite{Capogna-Bi2212}. For a comparison of the same measurements in samples having different masses, it is quite convenient to plot the inverse-flipping ratio $\rm 1/FR(T)$ where the magnetic signal should appear on top of the inverse of the bare flipping ratio, $\rm 1/FR^o(T)$ \cite{Baledent-YBCO}. Fig.~\ref{Fig3} compares 
$\rm 1/FR(T)$ in the 3 different samples at 2 different Bragg spots. The weak temperature dependence of $\rm 1/FR^o(T)$ is determined by an extra measurement at the Bragg peak (2,0,0), i.e at large $|\bf{Q}|$ where the magnetic signal becomes vanishingly small and can be ignored. In contrast, the inverse flipping ratio at {\bf Q}=(1,0,1) shows an enhancement upon cooling down (Fig.~\ref{Fig3}): such an enhancement highlights the IUC magnetic order,  given by $\rm [ 1/FR(T)- 1/FR^o(T)]$, that develops below $\rm T_{mag} \sim$ 230 K in sample UD-85 and below $\rm T_{mag} \sim$ 170 K in sample OD-87. At larger hole doping, in sample OD-70, no magnetic signal can be detected in the normal state. It is worth mentionning that we observe in Bi2212 a signal around optimal doping that was not possible to detect previously in Y123 \cite{Fauque} or 
Hg1201\cite{Li-Nature}. The nuclear Bragg (1,0,1) intensity is actually weaker in Bi2212 allowing to observe smaller magnetic moments.

\begin{figure}[t]
\includegraphics[width=7cm,height=6cm,angle=0]{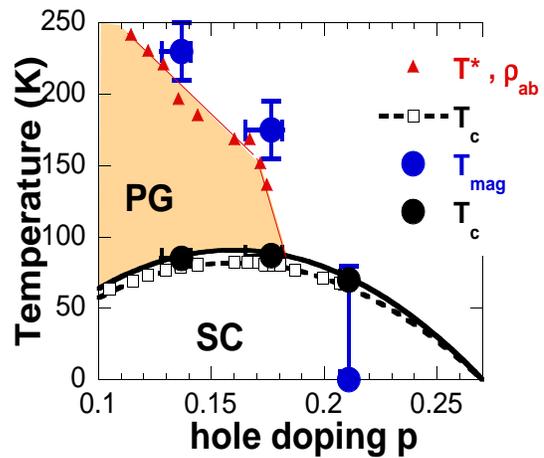}
\caption {
(color online) Hole doping dependence of the superconducting critical temperature $\rm T_c$ (Black dots) and magnetic ordering temperature $\rm T_{mag}$ (blue dots) corresponding to the large single crystals used for the polarized neutron scattering study. The data are compared with the temperature dependence of $\rm T_c$ (open squares) and the pseudo-gap temperature $\rm T^{\star}$ (red triangles) from ab-resistivity measurements on thin films \cite{Raffy}.
}
\label{Fig4}
\end{figure}

Fig.~\ref{Fig4} summarizes the variation of the magnetic ordering temperature $\rm T_{mag}$ as a function of the hole doping p, given by the empirical relationship: $\rm T_c(p) / T_c^{max}=1-82.6(p-0.16)^2$, \cite{Hole-doping} with $\rm T_c^{max}$ set to 91$\pm$2 K. $\rm T_{mag}$ decreases continuously upon increasing p and is likely to vanish upon approaching a critical hole level of $\sim$0.19, the end point of the PG state according to electronic specific heat measurements \cite{Loram}. $\rm T_{mag} (p)$ can be compared with $\rm T^{\star} (p)$ obtained from in-plane resistivity measurements \cite{Raffy}. At this step, it is worth pointing out that in-plane resistivity measurements are performed on thin films and $\rm T_c^{max}$ does not exceed 82 K \cite{Raffy}. One should therefore be particularly careful when comparing large single crystals and thin films, for which crystal growth techniques and/or annealing conditions are different. The comparison of $\rm T_{mag} (p)$ with $\rm T^{\star} (p)$, reported in Fig~\ref{Fig4}, shows that they both exhibit a similar evolution as a function of hole doping. In particular, $\rm T^{\star} (p)$ does not seem to decrease linearly with increasing p, but rather displays a shoulder close to optimal doping (p=0.16).  That persistence well above $\rm T_c$ near optimal doping is typical of the phenomenology of the PG state in Bi2212. This qualitative comparison suggests that the observed magnetic order and the PG state are likely bound to each other.

In the absence of a common theory for the pseudogap to guide data analysis, each experimental technique has developped their own definition to determine a temperature $\rm T^{\star}$.  As a result, values of $\rm T^{\star}$ may vary from one technique to another. It is therefore more meaningfull to consider the decay rate of $\rm T^{\star}$ upon increasing hole doping. According to ARPES \cite{Ding}, Knight shift NMR \cite{Ishida} and tunneling \cite{STM} measurements, the decay rate of $\rm T^{\star}$ and the PG energy $\rm \Delta_{PG}$ is about 1.5-1.7 from the weakly UD regime ($\rm T_c$=82$\pm$ 3 K) to the weakly OD regime ($\rm T_c$=85$\pm$ 3 K). Consistently, $\rm T_{mag}$ and the ordered magnetic moment at 100 K ($\rm |{\bf M}|= \sqrt{I_{mag}}$) decay by factors $\sim$ 1.4 and $\sim$ 1.7, respectively (Fig.~\ref{Fig3}). 

The observed 3D magnetic order does not appear to change upon structural properties of each hole-doped cuprate. It does not depend neither on the number of $\rm CuO_2$ planes per unit cell nor on the nature of stacking of Cu sites along the $\rm {\bf c}$ axis. Indeed, Bi2212 has a body centered structure, at variance with Y123. Hg1201 has a simple tetragonal structure, Y123 an  orthorhombic one along the Cu-O bonds as well as Bi2212 but along diagonals. In all cases, the observed order preserves the crystal lattice translation invariance, corresponding to an IUC-(antiferro)magnetic order\cite{CC-review}. 

 The existence of a magnetic order signals shows that time reversal symmetry is broken. This property in the PG state of Bi2212 was first inferred from the observation of dichroic effect in ARPES measurements \cite{Kaminski-Dichroism}, but this measurement has been the subject to a long standing controversy (see {\it e.g.} \cite{arpiainen} and related discussions). Keeping in mind that ARPES measurements are performed on thin films \cite{Kaminski-Dichroism} and polarized neutron measurements on large single crystals, the onset of time reversal breaking symmetry found in both types of measurements for samples UD-85 is in a very good agreement ($\sim$200K for ARPES and $\sim$230 K for polarized neutron diffraction). The doping dependence obtained in both types of experiments also match each other quite well. 

Symmetry breaking within the PG state is now corroborated on various cuprates by different techniques \cite{Kapitulnik} in addition to polarized neutron diffraction and the circularly polarized ARPES measurements. In monolayer $\rm Pb_{0.55}Bi_{1.5}Sr_{1.6}La_{0.4}Cu_{6+\delta}$\cite{Pb-Bi2201}, ARPES, polar Kerr effect and time-resolved reflectivity measurements performed on the same single crystal demonstrate that a phase transition breaking time reversal symmetry takes place at $\rm T^{\star}$. 
To further characterize the actual order parameter, it is interesting to notice that the existence of an IUC order associated with the PG state in Bi2212 system is also supported by recent analyses \cite{Lawler} of scanning tunneling microscope (STM) images. An electronic nematic order has been observed as the $\rm 90^{o}$ rotational symmetry is broken in the $\rm CuO_2$ unit cell, yielding distinct electronic densities on oxygen sites along a and b crystal axis. In principle, this IUC order differs from the CC-loop order. However, the mean-field analysis of the different IUC-ordering possibilities in the 3-band Emery model indicate that the electronic nematic order and the CC-loop order could actually coexist \cite{Fischer}.

Finally, the concomittant observations in Bi2212 of a dichroic effect at the antinodal wave vectors and a magnetic signal on Bragg reflections (1,0,L) point towards a PG order parameter breaking inversion symmetry \cite{Varma-Simon,Agtenberg} with the result of dismissing the model of spin moments on oxygen atoms \cite{Fauque} and promoting the loop current CC-$\rm \theta_{II}$ phase \cite{Varma97}. Recent observations of two nearly dispersionless magnetic excitations in Hg1201 by polarized inelastic neutron scattering \cite{Li-INS} give actually further evidence for the presence of a such discrete order parameter \cite{Varma-INS}.

{\it Acknowledgments}: We wish to thank L. Ammor, J. C. Davis, M. Greven, Y. Gallais, M-H. Julien, Eun-Ah Kim, B. Leridon, A. Ruyter and C.M. Varma for valuable discussions on various aspects related to this work.



\begin{thebibliography}{100}




\bibitem{Kaminski-Dichroism} A. Kaminski {\it et al.},  Nature {\bf 416}, 610 (2002).

\bibitem{CC-review} P. Bourges, and Y. Sidis, C. R. Physique, {\bf 12}, 461, (2011).

\bibitem{Kapitulnik} J. Xia {\it et al.}, Phys. Rev. Lett {\bf 100} 127002 (2008).

\bibitem{Lawler} M. J. Lawler {\it et al.}, Nature {\bf 466} 347 (2010).

\bibitem{Pb-Bi2201} R.-H. He {\it et al.}, Science {\bf 231}, 1579 (2011).



\bibitem{Varma97}  C. M. Varma, Phys. Rev. B {\bf 73}, 155113 (2006).

\bibitem{Varma-Simon} M.E. Simon, and C. M. Varma, Phys. Rev. Lett. {\bf 89}, 270003, (2002).


\bibitem{Greiter07} R. Thomale, and  M. Greiter, Phys. Rev. B {\bf 77}, 094511 (2008).

\bibitem{Weber} C. Weber {\it et al.}, Phys. Rev. Lett. {\bf 102} 017005 (2009).

\bibitem{Gabay08}  P. Chudzinski {\it et al.} , Phys. Rev. B {\bf 78}, 075124 (2008).



\bibitem{Fauque} B. Fauqu\'e {\it et al.},  Phys. Rev. Lett. {\bf 96} 197001 (2006).

\bibitem{Sidis} Y. Sidis {\it et al.}, Physica B, {\bf 397}, 1 (2007).

\bibitem {Mook} H. A. Mook {\it et al.}, Phys. Rev. B {\bf 78} 020506 (2008).

\bibitem{Baledent-YBCO} V. Bal\'edent {\it et al.}, Phys. Rev. B {\bf 83}, 104504 (2011).

\bibitem{Li-Nature} Y. Li {\it et al.}, Nature 445 {\bf 372} (2008).

\bibitem{Li-PRB} Y. Li {\it et al.}, Phys. Rev. B {\bf 84}, 224508 (2011).

\bibitem{Baledent-LSCO} V. Bal\'edent {\it et al.}, Phys. Rev. Lett. {\bf 105}, 027004 (2010).




\bibitem{Lederer} S. Lederer, and S. A. Kivelson, Phys. Rev. B {\bf 85}, 155130 (2012).


\bibitem{Strassle-2} S. Str\"assle {\it et al.}, Phys. Rev. Lett. {\bf 106},  1 (2011).


\bibitem{Leridon} B. Leridon {\it et al.}, Europhys. Lett., {\bf 87}   17011 (2009).



\bibitem{Gronsleth} M. S. Gronsleth {\it et al.},  Phys. Rev. B {\bf 79} 094506 (2009).

\bibitem{Aji} V. Aji {\it et al.}, Phys. Rev. B {\bf 78}, 094421 (2008).


\bibitem{Orenstein} J. Orenstein, Phys. Rev. Lett. {\bf 107}, 067002 (2011).


\bibitem{Scagnoli} V. Scagnoli {\it et al.}, Science {\bf 332}, 696 (2011).



\bibitem{Renner-RMP} S. Fisher{\it et al.}, Rev. Mod. Phys. {\bf 79}, 353 (2007).

\bibitem{Damascelli-RMP} A. Damascelli {\it et al.}, Rev. Mod. Phys. {\bf 75} 473 (2003).


\bibitem{Agtenberg} R. P. Kaur, and D. F. Agterberg, Phys. Rev. B {\bf 68}, 100506 (2003).









\bibitem{DeAlmeida-sample} S. De Almeida-Didry {\it et al.}, J. Crystal Growth , {\bf 312} ,  466-470   (2010).


\bibitem{Etrillard} J. Etrillard {\it et al.}, Phys. Rev. B {\bf 62}, 150 (2000).

\bibitem{Fauque-Bi2212} B. Fauqu\'e {\it et al.}, Phys. Rev. B, {\bf 76}, 214512 (2007).

\bibitem{Capogna-Bi2212} L. Capogna {\it et al.}, Phys. Rev. B {\bf 75}, 060502 (2007).

\bibitem{Hole-doping} J.L. Tallon  {\it et al.}, Phys. Rev. B {\bf 68}, 180501 (2003).

\bibitem{Loram} J. W. Loram {\it et al.}, Physica C {\bf 341-348}, 831-833 (2000).

\bibitem{Raffy} H. Raffy {\it et al.}, Physica C {\bf 460-462},  851 (2007); Z. Konstantinovic {\it et al.}, Physica C {\bf 259-261},  567 (1999). 

\bibitem{Ding} H. Ding {\it et al.}, Nature {\bf 382}, 51 (1996).

\bibitem{Ishida} K. Ishida {\it et al.}, Phys. Rev. B {\bf 58}, R5960 (1998).

\bibitem{STM} R. M. Dipasupil {\it et al.}, J. Phys. Soc. Jpn {\bf 71}, 1535 (2002).

\bibitem{arpiainen} V. Arpiainen {\it et al.}, Phys. Rev. Lett. {\bf 103}, 067005 (2009). 




\bibitem{Fischer} M. H. Fischer, and E.-A. Kim, Phys. Rev. B {\bf 84}, 144502 (2011).





\bibitem{Li-INS} Y. Li {\it et al.}, Nature {\bf 468}, 283 (2010); Nat. Phys.  {\bf 8}, 404 (2012).

\bibitem{Varma-INS} Yan He, and C.M. Varma, Phys. Rev. Lett. {\bf 106}, 147001 (2011).

\end{thebibliography}
\end{document}